\documentclass[]{article}

\usepackage{color}
\usepackage{array}
\usepackage{fullpage}
\usepackage{graphicx}
\usepackage{epstopdf}
\usepackage{wrapfig}
\usepackage{colortbl}
\usepackage{amsmath}
\usepackage{amssymb}
\usepackage{authblk}

\begin{document}

\title{Dynamic Graph Metrics: Tutorial, Toolbox, and Tale}
\author[1]{Ann E. Sizemore}
\author[1,2,*]{Danielle S. Bassett}

\affil[1]{Department of Bioengineering, University of Pennsylvania, Philadelphia, PA, 19104}
\affil[2]{Department of Electrical and Systems Engineering, University of Pennsylvania, Philadelphia, PA, 19104}
\affil[*]{To whom correspondence should be addressed: dsb@seas.upenn.edu}

\date{\today}

\maketitle

\begin{abstract}
The central nervous system is composed of many individual units -- from cells to areas -- that are connected with one another in a complex pattern of functional interactions that supports perception, action, and cognition. One natural and parsimonious representation of such a system is a graph in which nodes (units) are connected by edges (interactions). While applicable across spatiotemporal scales, species, and cohorts, the traditional graph approach is unable to address the complexity of time-varying connectivity patterns that may be critically important for an understanding of emotional and cognitive state, task-switching, adaptation and development, or aging and disease progression. Here we survey a set of tools from applied mathematics that offer measures to characterize dynamic graphs. Along with this survey, we offer suggestions for visualization and a publicaly-available MATLAB toolbox to facilitate the application of these metrics to existing or yet-to-be acquired neuroimaging data. We illustrate the toolbox by applying it to a previously published data set of time-varying functional graphs, but note that the tools can also be applied to time-varying structural graphs or to other sorts of relational data entirely. Our aim is to provide the neuroimaging community with a useful set of tools, and an intuition regarding how to use them, for addressing emerging questions that hinge on accurate and creative analyses of dynamic graphs.
\end{abstract}

\maketitle

\twocolumn
\section*{Introduction}
\label{sec0}

The mammalian brain is a complex system, composed of many individual units (cells, neural ensembles, voxels, or areas) that are intricately connected with one another \cite{bassett2017network}. Understanding this system requires complementary studies from both reductionistic and holistic perspectives \cite{bassett2011understanding}. Reductionistic approaches are critically necessary to understand the structure and function of individual units, while holistic approaches are critically necessary to understand how those individual units function in the context of others. Historically, constructing and testing hypotheses regarding systems or subsystems of interconnected units has proven challenging, in large part due to a dearth of appropriate theories and associated computational tools \cite{newman2011complex}. Recent developments in network science \cite{newman2010networks} provide a wealth of potentially useful solutions to this problem by representing complex systems as graphs in which nodes (units) are connected by edges (interactions). This network representation forms a natural mathematical framework in which to couch holistic inquiries into the nature of the brain \cite{bullmore2011brain,sporns2015cerebral} and can be flexibly applied to neural data collected across spatial and temporal scales \cite{betzel2016multi}, across species \cite{van2016comparative}, and across cohorts \cite{stam2014modern,fornito2015connectomics}.

One canonical form of interest to neuroscientists is the functional graph in which cells, neural ensembles, voxels, or areas are connected to one another by estimates of their functional (rather than structural) interactions. At the neuronal scale, a functional edge might be an estimate of similarity in firing patterns \cite{feldt2009functional}, while at the large scale, it might be an estimate of similarity in BOLD time series \cite{achard2006resilient} or ECOG signals \cite{kramer2011emergence,burns2014network,khambhati2016virtual}. Irrespective of spatial scale, when considering how to build a functional network representation from neural data, one is faced with the natural question of whether a single representation will suffice, or whether an ensemble of representations is required. Early but very important work in this field focused on constructing a single representation \cite{stam2007small,devicofallani2007cortical,meunier2009age,bassett2006adaptive}, in which an edge summarized functional interactions between two neural units over a fixed time period. However, this approach is incompatible with the emerging interests in understanding the network dynamics -- and not just its structure -- that support cognition \cite{medaglia2015cognitive}. Indeed, querying (i) fluctuations in an animal's emotional or cognitive state \cite{betzel2017positive,fornito2012competitive,shine2016temporal}, (ii) the manner in which an animal transitions between tasks \cite{braun2015dynamic,ueltzhoffer2015stochastic}, or (iii) the variations in functional network architecture that are characteristic of perception and processing \cite{chai2016functional}, learning \cite{heitger2012motor,mantzaris2013dynamic}, development \cite{fair2009functional,gu2015emergence}, aging \cite{meunier2009age,betzel2014changes}, or disease progression \cite{raj2015network} all require an assessment of a network's dynamics.

The last several years have seen a proliferation of approaches to quantitatively describe time-varying patterns of functional connectivity \cite{hutchison2013dynamic,calhoun2014chronnectome,betzel2016multi}. One set of powerful tools comes from engineering approaches including independent components analysis, machine learning, and causal inference, while another set comes more from the field of pure and applied mathematics and specifically graph theory. In some ways the distinction between these two types of approaches is reminiscent of the distinction between model-free \emph{versus} model-based learning \cite{daw2014algorithmic}: graph theory-based approaches assume a formal graph model of the data, while other approaches seek to learn a model directly from the data. Though each approach has its benefits, we focus our exposition here on the graph-based approach due to the recent explosion of tools developed by the applied mathematics community to study dynamics graphs -- also called \emph{temporal networks} \cite{holme2012temporal}. These advances form a potentially powerful toolset for the contemporary neuroscientist, paving the way to more sophisticated approaches to data analysis and to hypothesis development. 

Here we offer a didactic piece that describes dynamic graphs, discusses how to visualize them, surveys dynamic graph measures, and demonstrates their application to a previously published neuroimaging data set. We devote slightly less real estate to tools that have already been applied to neuroimaging data, and slightly more real estate to tools that have not yet been applied in this area. Along with this exposition, we offer a publically-available MATLAB toolbox \cite{Sizemore2017} so that the reader can immediately apply these measures to their own data to address their own hypotheses. The piece can be thought of as a mathematical reference and does not attempt to provide new neurophysiological insights (we leave the latter to future forays by interested readers). Finally, we note that although we illustrate these tools in the context of time-varying functional brain graphs, the toolset is flexible and can be applied to questions regarding time-varying structural or morphometric graphs as well.

The remainder of this paper is structured as follows. First, we describe different ways of visualizing dynamic graphs and discuss the advantages and disadvantages of each. Next, we discuss how to encode a dynamic graph and then describe several basic dynamic graph notions and measures including time-respecting paths, latency, and centrality. We then move on to a discussion of null models and additional measures including temporal small-worldness and dynamic modular structure. Finally, we outline a few natural scenarios in which dynamic graphs could be constructed to address hypotheses regarding brain structure and function as well as the neurophysiological mechanisms of behavior and disease.

\section*{Visualizing dynamic graphs}
\label{sec1}

\begin{figure}[h]
	\centering 
	
	\includegraphics[width = .9\linewidth]{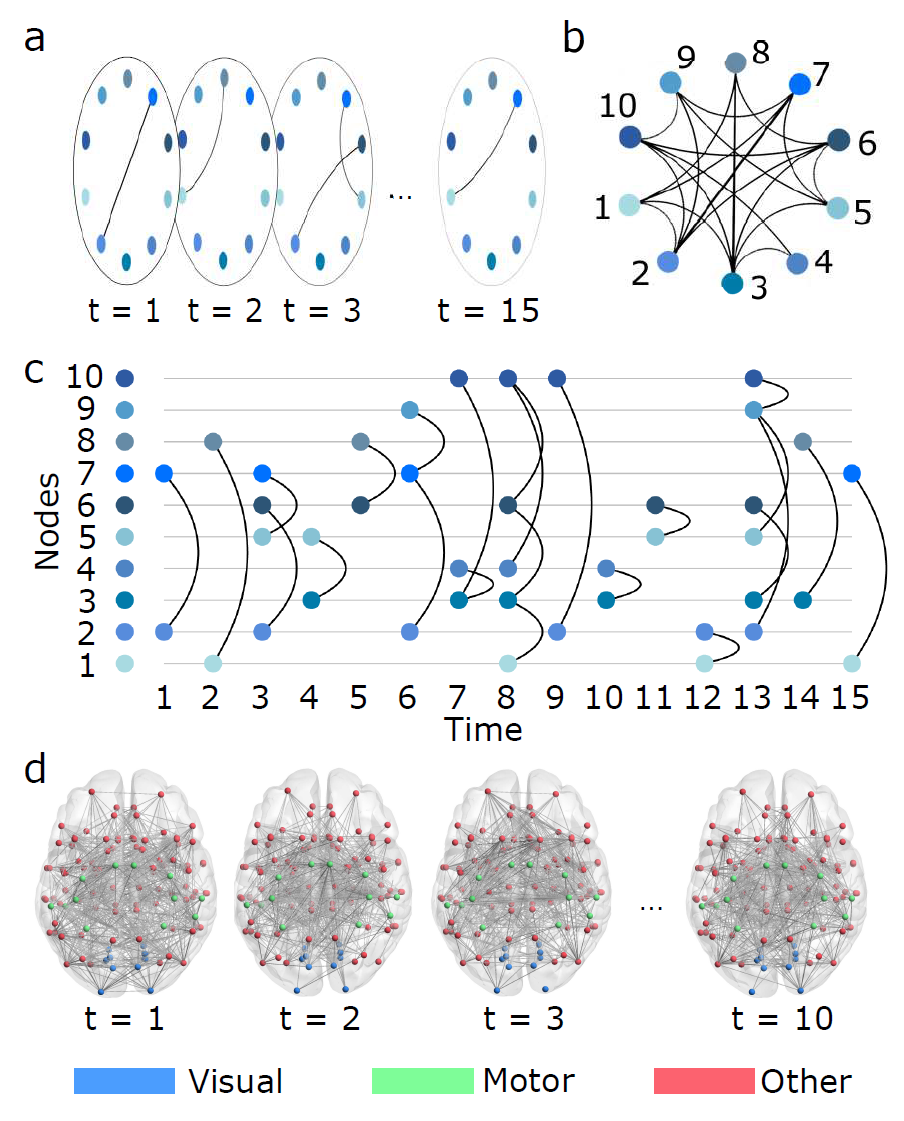}
	\caption{Visualizations of dynamic networks. \emph{(a)} Stacked static network representation of a dynamic network on ten nodes. \emph{(b)} Time-aggregated graph of dynamic network in \emph{(a)}. Any two nodes that are connected at any time in \emph{(a)} are connected in this graph. \emph{(c)} Visualization of network in \emph{(a)} as contacts across time. \emph{(d)} Dynamic network of one individual during a motor learning task \cite{bassett2015learning}. Green regions correspond to a functional module composed of motor areas, blue regions correspond to a functional module composed of visual regions, and red regions correspond to areas that were not in either the motor or visual module. }
	\label{fig:1}
	
\end{figure}

Given data as a dynamic graph, a first inclination is to find a way to visualize the information. For simplicity we will assume this graph is undirected and binary, and that edges can exist at any of some finite number of timepoints. We may naturally imagine viewing the dynamic network as a movie where edges and nodes come in and out of view. Since a movie is not always feasible (for example in research papers), we might look to study the frames, or snapshots of the dynamic network at each timepoint, as seen in Fig.~\ref{fig:1}a. While this approach certainly captures information from the time dimension, it becomes less helpful as the number of timepoints increases. Particularly for sparse networks, it may be more useful to visualize a collapsed, static graph (Fig.~\ref{fig:1}b), specifically the \emph{time-aggregated graph}, where edges exist between two nodes if they are connected at any point in the dynamic network \cite{holme2012temporal}. Note this time-aggregated graph is created from a dynamic network describing the data, instead of a more traditional approach of creating a single graph from multiple time points which averages out the dynamics. Here edge weights could be assigned by the time of edge appearance or frequency of edges within the dynamic network. We then gain a more succinct and holistic view of the dynamic network, yet lose comprehension of temporal structure. As a third option, we could also explicitly visualize the time dimension by plotting the dynamic network as a sequence of edges or \emph{contacts} over time, giving a circuit board-like view (Fig.~\ref{fig:1}c). This approach is optimal for small, sparse dynamic networks, though quickly becomes overwhelming as the number of nodes and contacts grows. More methods for visualization exist, but, for the optimal representation, one should consider the size and density of the given data. 

One example of data suitable for a dynamic network encoding is functional magnetic resonance imaging (fMRI) data. Here we illustrate dynamic graph approaches using fMRI scans collected as individuals learned to play a sequence of finger movements \cite{bassett2015learning}. The time-dependent levels of neural activity from $N=112$ cortical and subcortical brain regions were estimated from indirect measurements of blood-oxygen-level dependent (BOLD) signal collected over ten time windows in each of four training sessions. Functional connectivity between brain regions was estimated with a magnitude squared coherence of wavelet coefficients \cite{sun2004measuring}, resulting in an $N \times N$ coherence matrix for each time window in each training session. While prior studies have examined these coherences matrices as fully weighted graphs, for the didactic purposes of this tutorial we simplify the data by binarizing the dynamic network, keeping only the top 10\% of entries in each coherence matrix. Prior analyses of this data provided insight into how individuals learned on short (within one session) and long (across sessions) timescales \cite{bassett2015learning,bassett2013robust,bassett2013task,wymbs2015human}. For this type of data, a fourth type of visualization is available, namely visualizations that place nodes in their true anatomical locations and draw lines between connected nodes. In Fig.~\ref{fig:1}d we show this exact type of visualization for a dynamic network from one individual from the first session as a sequence of brain graphs. We choose to color nodes according to prior observations on these same data: the presence of two groups of densely connected brain regions, a group of motor regions, which we color in green, and a group of visual regions which we color blue \cite{bassett2015learning}. All other regions, shown in red, were not found to have any particular allegiance to either module.

\section*{Basic measures}
\label{sec2}

In this section, we discuss how to encode a dynamic graph and then describe several basic dynamic graph notions and measures including time-respecting paths, latency, and centrality. 

\subsection*{Encoding data as a dynamic graph}

Data to be analyzed as a dynamic graph may arrive in different formats, including a sequence of matrices or a list of edges and times. Thus, before we begin any calculations, we might wish to transform the information into a standard -- and efficiently stored -- object. For a static graph, this is simply $G = (V,E)$, where the graph $G$ is defined by a set of vertices $V$ and edges $E:V\times V \rightarrow \mathbb{R}$. For a dynamic graph, we could record $G_{0}, G_{1}, \dots G_{T}$ for each time-point $t = 0,1,\dots,T$, but it is more memory-efficient to record instead the list of contacts and the time at which these occur. For a dynamic network, a \emph{contact} is a triple $(i,j,t)$ indicating the existence of an edge between nodes $i$ and $j$ (or from node $i$ to node $j$ in the directed case) at time $t$. Then the set of contacts in our dynamic network is called the \emph{contact sequence} and this is how we will record and work with our dynamic network. Note this can be expanded to include more information, such as edge weight or time delay required to traverse the edge, by defining contacts to be tuples $(i,j,t,w_1,w_2,\dots,w_k)$ for the additional measures $w_m$. 

With our dynamic network efficiently encoded, we can begin asking questions about its structure and evolution. At the level of individual nodes, many measures are intuitively generalizable as a function of time, for example the clustering coefficient \cite{saramaki2007generalizations}. Similarly, we can also track global measures -- such as efficiency \cite{latora2001efficient} -- across time as well. However, not all measures can (or should) be simply extended in this way, because it ignores the evolution of the network from one timepoint to the next. Indeed, by ignoring the temporal dependencies between consecutive graphs, one is assuming that each observation is independent from the others; not only does this lead to inaccuracies in statistical testing and inference \cite{lebre2010statistical,bassett2011dynamic}, but it also means that the investigator is unable to identify temporal motifs (analogous to topological motifs studied in static graphs \cite{shoval2010snapshot,sporns2004motifs}) -- characteristic changes in or reconfiguration of the network that may happen with some unexpectedly high or low frequency \cite{kovanen2013temporal,xuan2015temporal}. Dynamic graph metrics address these limitations by explicitly accounting for the fact that the set of graphs is ordered in time. Due to their enhanced statistical rigor, we focus solely on dynamic graph metrics in this review.

\subsection*{Time-respecting paths}

\begin{figure}
	\centering 
	
	\includegraphics[width = .9\linewidth]{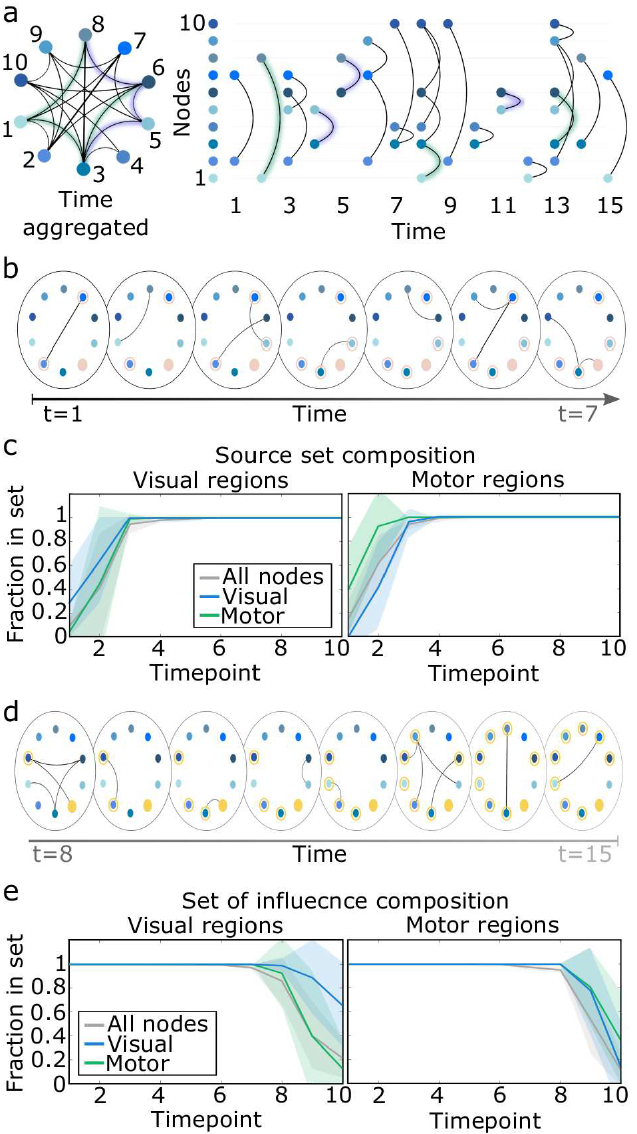}
	\caption{Time respecting paths. \emph{(a)} (Left) Time aggregated network from Fig.~\ref{fig:1}b with green and blue paths highlighted. (Right) Contact sequence plot from Fig.~\ref{fig:1}c with green and blue paths highlighted. \emph{(b)} The source set of the peach node indicated with a peach ring.  \emph{(c)} Composition of the source set of nodes from the visual (left) and motor (right) modules of our example empirical fMRI data set, depicted across time. The gray line indicates the fraction of all nodes in the source set, while the blue and green lines represent the fraction of the visual and motor nodes within the source set, respectively. \emph{(d)} Illustration of the set of influence ($t-8$) of the gold node. Nodes within this set indicated with a gold ring at the time at which they can first be reached by the gold node. \emph{(e)} Composition of the set of influence calculated from nodes within the visual (left) and motor (right) groups. As in \emph{(c)}, the fraction of all regions (gray), visual regions (blue), and motor regions (green) are plotted against time. Solid lines in \emph{(c)} and \emph{(e)} mark the average over subjects and trials, and shaded regions represent two standard deviations from this average.}
	\label{fig:2}
	
\end{figure}

Paths and connectivity within a static graph can be indicative of trajectories of information spreading. In a dynamic network, the time dimension induces an additional restriction on connectivity. For example, in Fig.~\ref{fig:2}a (left), we see the time-aggregated graph of our model dynamic network from Fig~\ref{fig:1}. The edges highlighted in green and purple connect as two valid paths in this static network. Yet, we see in Fig.~\ref{fig:2}a (right) when looking at the sequence of contacts that the purple path is not a valid path in the dynamic network. Said another way, if information was sent from node 3, it could not reach node 8 via this sequence of contacts. Conversely, information from node 8 could reach node 6 by following the sequence of green contacts. Such a collection of contacts is called a \emph{time-respecting path}. Precisely, a time-respecting path is a sequence of contacts $(n_0,n_1,t_0), (n_1,n_2,t_1), \dots, (n_{k-1},n_k,t_{k-1})$ such that $t_i < t_{i+1}$ for all $i = 0,..., k-2$. Defined in this way, these time-respecting paths must agree with the ``arrow of time," thereby making them particularly useful for the study of information flow in dynamic networks. 

The notion of a time-respecting path provides important intuitions regarding the similarities and differences between static and dynamic graphs. Returning to the model dynamic network in Fig~\ref{fig:2}a, note that we have a time-respecting path from node 3 to node 6 and from node 8 to node 6, yet no path exists from node 3 to node 8. Unlike in static graphs, time-respecting paths in dynamic networks are not required to be transitive. That is, if a path from node $a$ to node $b$ exists and a path from node $b$ to node $c$ exists, this does not imply the existence of a path from node $a$ to node $c$. Thus, when studying systems from both static and dynamic perspectives, it is important to maintain accuracy in interpreting the potential utility of paths for information transmission.

The notion of time-respecting paths can also allow us to study the reachability of a node, which may be an important indicator of its function. For example, a brain region that can be reached from many other regions via time-respecting paths may have a significant role in information integration. Then, the set of nodes that can reach our node of interest also becomes a key feature. For example, in Fig.~\ref{fig:2}b, we ask which nodes connect to the peach node through time-respecting paths by $t=7$. In other words, at $t=7$, which nodes could be the source of the peach node's view of the system? This is called the \emph{source set} of the peach node, and those within this set are circled in peach once they participate in a time-respecting path to the peach node. We have chosen a specific timepoint in this example, but one could record this at each point in time. Then for each node, the size and composition of the source set could inform that node's function. In our example empirical fMRI network, throughout one session we calculate the size and makeup of the source set for nodes in the visual and motor groups (Fig.~\ref{fig:2}c). We see that a larger fraction of the visual group (blue) than the motor group (green) is part of the source set for visual regions and conversely for the motor regions. This intuitively makes sense, as we might expect visual regions to be contacted by many visual regions and \emph{vice versa} for the motor regions.

We could now invert the source set concept and look forward instead of backward in time for a node. Instead of who connects to a node, we can ask who this node can influence? If the gold node in Fig.~\ref{fig:2}d learns something new just before $t=8$, we can look forward in time and find the other nodes with which the gold node can share this new information. We call this the \emph{set of influence}: the collection of nodes reachable via time-respecting paths beginning no earlier than a given time $t$, which we illustrate as all nodes circled in gold at the final timepoint in Fig.~\ref{fig:2}d. Similar to the source set, we can calculate the number and makeup of this set as we vary $t$. In our example empirical fMRI data, we see that, as time increases, the visual regions influence many of the visual and motor regions, while the motor regions are more often connecting to strictly motor regions. With a deferential nod to notions from astrophysics, Holme and Samaraki describe these two sets, the source set and the set of influence for a node at a particular time, together as ``light cones" which either could have affected the current state of the node or will be affected by the current state of the node \cite{holme2012temporal}.

\subsection*{Latency and centrality}

The notions in the previous section provided us with information about the connectivity of nodes in a dynamic graph. Next we turn to questions related to the speed at which those nodes might communicate. In a static network, the number of edges within a path defines the path length, while in a dynamic network we can additionally record the duration of the path. We call the difference in time between the first and last contact the \emph{temporal path length} \cite{pan2011path}. For particularly efficient systems, one might expect information to travel along the shortest -- or more precisely, the fastest -- path within the dynamic network. Then, the distance between two nodes can be measured with temporal path length. We use the term \emph{latency} (or temporal distance \cite{pan2011path}) of nodes $i$ and $j$ to refer to the shortest time it takes to move from node $i$ to node $j$. 

\begin{figure}
	\centering 
	
	\includegraphics[width = .9\linewidth]{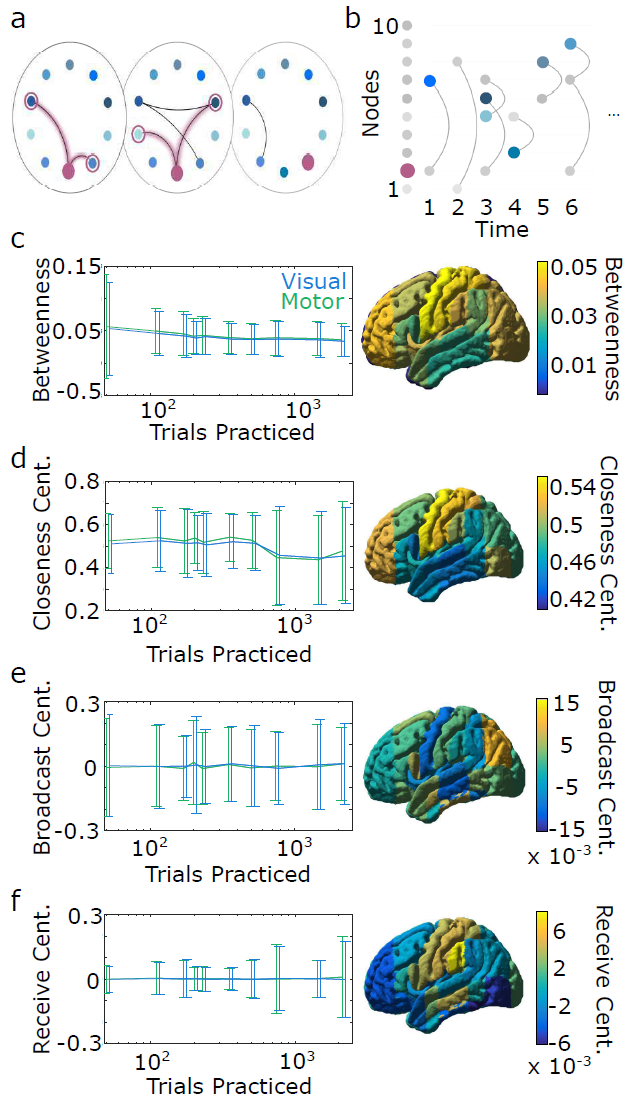}
	\caption{Centrality in dynamic networks. \emph{(a)} Time window of the model network shown in Fig.~\ref{fig:1}a--c highlighting the fastest paths that pass through the maroon node, and therefore affect its betweenness centrality. \emph{(b)} Schematic of closeness centrality for the maroon node in the model network. Closeness centrality measures the speed at which a node can reach all others: the time at which other nodes are first reached by node 2 determines its closeness centrality. Nodes are shown in color at the earliest time they are reached by node 2. \emph{(c--f)} An illustration of the notions of centrality for our example empirical fMRI data shown in Fig.~\ref{fig:1}d. \emph{(c)} (Left) Betweenness centrality for visual (blue) and motor (green) regions as a function of the number of trials practiced. (Right) Averaged betweenness centrality scores across trials practiced for each brain region. \emph{(d)} Closeness centrality for visual and motor regions during learning (left), and (right) averaged over number of trials as in \emph{(c)}. \emph{(e)} Broadcast centrality for visual and motor regions during learning (left), and the same values now averaged over all trials (right). \emph{(f)} Receive centrality for visual and motor regions during learning (left), and the same values now averaged over all trials (right). Error bars indicate two standard deviations from the mean over subjects and trials practiced.}
	\label{fig:3}
	
\end{figure}

Defining latency as the measure of shortest distance (note now in a temporal sense) allows us to extend notions of centrality to dynamic networks. Recall that in a static network, the betweenness centrality of a node can be defined as the fraction of shortest paths passing through that node, or 

\begin{equation} \label{eq:CB}
C_B(i) = \sum_{i\neq j \neq k} \frac{\sigma_{j,k}(i)}{\sigma_{j,k}}\\,
\end{equation}

\noindent with $\sigma_{j,k}$ being the number of shortest paths between nodes $j$ and $k$ and $\sigma_{j,k}(i)$ being the number of shortest paths passing through node $i$ \cite{easley2010networks,jackson2010social}.  Using the definition of temporal path length, we can compute the same notion but for dynamic networks \cite{tang2010analysing} by swapping the shortest path for the fastest path within a specified time window. In this way, we see the \emph{temporal betweenness centrality} can be written as
 
 \begin{equation} \label{eq:tempCB}
 C_B(i,t) = \sum_{i\neq j \neq k}\frac{\sigma_{j,k}(i,t)}{\sigma_{j,k}(t)}\\,
 \end{equation}
 
\noindent  if we let $\sigma_{j,k}(t)$ be the number of fastest paths from node $j$ to node $k$ beginning no earlier than time $t$. In Figure~\ref{fig:3}a, we illustrate these concepts for the toy dynamic graph shown in Figure~\ref{fig:1}a--c. Specifically, we show dynamic network features used in the calculation of betweenness centrality for a single node in the graph: highlighted nodes and edges participate in fastest paths involving the node of interest. An interesting alternative definition of temporal betweenness centrality swaps the fastest time-respecting paths for the shortest topological time-respecting paths: those with the fewest hops throughout the dynamic network \cite{holme2012temporal}.  

While quantifying and understanding the shortest paths between nodes could be quite interesting, we might also wish to measure how far all other nodes are from a node of interest. In static graphs, we know this as closeness centrality, defined as

\begin{equation} \label{eq:CC}
C_C(i) = \frac{N-1}{\sum_{j \neq i} d(i,j)}\\,
\end{equation}

\noindent where $d(i,j)$ is the distance (path length) between node $i$ and node $j$, and $N$ the number of nodes in the network. When considering dynamic graphs, we could simply swap $d(i,j)$ here for the latency between node $i$ and node $j$, which takes into account the whole dynamic network. But if information is given to node $i$ at some time $t$, it might be more relevant to measure how fast this information from node $i$ will reach the rest of the nodes. For this reason, we define the \emph{forward latency} $\tau(i,j,t)$ as the time it takes to reach node $j$ from node $i$ via a time-respecting path beginning no earlier than $t$ \cite{pan2011path}. If node $i$ and node $j$ are disconnected, $\tau(i,j,t) = \infty$. Now we can substitute $\tau(i,j,t)$ for $d(i,j)$ in Eq~\ref{eq:CC} to recover the \emph{temporal closeness centrality}, 
 
 \begin{equation} \label{eq:tempCC1}
 C_C(i,t) = \frac{N-1}{\sum_{j \neq i} \tau(i,j,t)}\\,
 \end{equation}

\noindent for node $i$ and time $t$ \cite{wu2014path,nicosia2013graph,batagelj2016algebraic,kim2012temporal}. Because in practice we often observe disconnected nodes, we alter Eq.~\ref{eq:tempCC1} slightly by taking the mean of the inverse distance 

\begin{equation} \label{eq:tempCC2}
 C_C(i,t) = \frac{1}{N-1}\sum_{j \neq i} \frac{1}{\tau(i,j,t)}\\,
 \end{equation}

\noindent which allows us to account for disconnected nodes more cleanly \cite{pan2011path}.

While several other notions of centrality exist for temporal networks \cite{taylor2015eigenvector}, we will describe only two more in this review, chosen based on their theoretical relevance to neuroimaging data and neuroscientific hypotheses. Within a complex system such as the brain, we often simplify information pathways by assuming that only paths of shortest length or shortest time are essential. However, it is intuitively plausible that information can in fact follow any and all paths, but perhaps those of longer length are less critical than paths of shorter length. 

To formalize this idea, we can assign a weight $\alpha^k$ to paths of length $k$, $\alpha \in (0,1)$. This gives a richer perspective on how well node $i$ could potentially communicate with node $j$. Following \cite{mantzaris2013dynamic}, we compute the product of matrix resolvents

\begin{equation} \label{eq:broadcast1}
P := (I-\alpha A(1))(I-\alpha A(2))...(I-\alpha A(T))\\,
\end{equation}

\noindent for the binary matrices $A(1), A(2) \dots, A(T)$ encoding the binary temporal slices of the network at each timepoint. To avoid underflow and overflow, $P$ is normalized 

\begin{equation} \label{eq:broadcast2}
Q = \frac{P}{||P||_2}\\,
\end{equation}

\noindent so that the entry $Q_{i,j}$ describes the ability of node $i$ to communicate with node $j$ through paths of all lengths. Then we have the \emph{broadcast centrality} of node $i$, 

\begin{equation} \label{eq:broadcast3}
b(i) := \sum_{j=1}^N Q_{i,j}\\,
\end{equation}

\noindent and flipping the direction by summing over the rows we recover the \emph{receive centrality} of node $i$,

\begin{equation} \label{receive}
r(j) := \sum_{i=1}^N Q_{i,j} \hspace{2pc}\\
\end{equation}

\noindent describing the ability of all other nodes to communicate with node $j$. Together these two measures quantify how well nodes can reach and be reached by others along paths of all lengths.

Returning to our example empirical dynamic graph obtained from fMRI data, we observe the highest broadcast centrality in a broad swath of posterior parietal cortex extending to the posterior temporal fusiform cortex. By contrast, we observe the highest receive centrality in a broad swath of somatomotor and premotor cortex extending to the anterior supramarginal gyrus. Note that these anatomical distributions are complementary to but not redundant with the anatomical distributions of betweenness centrality and closeness centrality, which tend to display high values in frontal cortex and motor cortex. These differences are due to inherent differences in the underlying mathematical formulation: the broadcast and receive centrality capture the two sides of dynamic communicability \cite{grindrod2011communicability} and can be used to probe how individual brain regions distribute information across the network and across time.

 \section*{Null models and additional measures}
\label{sec3}
 
 \subsection*{Null models}

While summary statistics of dynamic networks offer insight into the temporal network structure, it is also critical to determine whether the architecture we observe differs significantly from that expected under an appropriate statistical null model. Addressing this question requires that we define and exercise dynamic network null models. For static graphs, common null models include the Erd\"{o}s-R\'{e}nyi random graph model \cite{erdos1960evolution}, the ring lattice \cite{watts1998collective}, and the configuration model \cite{newman2003structure}, to name a few. In principle, each of these static graph models can be extended to temporal graph models. However for simplicity, here we will focus only on the two most common dynamic network null models. 
 
\begin{figure}[t]
  	\centering 
  	
  	\includegraphics[width = .9\linewidth]{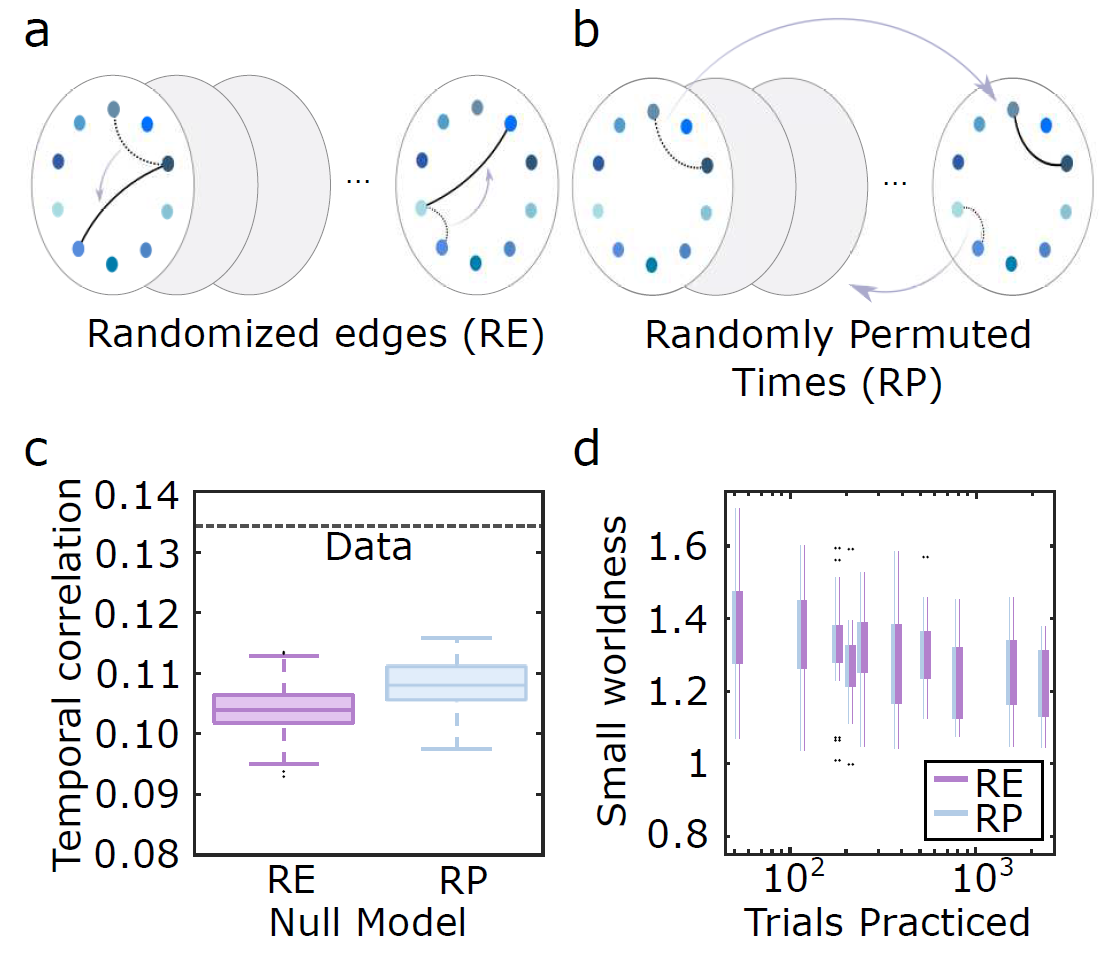}
  	\caption{Null models and their utility in measuring small-worldness in dynamic graphs. \emph{(a)} Schematic of the edge rewiring process for the \emph{randomized edges} (RE) model. \emph{(b)} Schematic of the \emph{randomly permuted times} (RP) model where contact times are permuted uniformly at random. \emph{(c)} Temporal correlation coefficients for one session of a participant in the study (black dashed line), and the 100 runs of the RE and RP model created from this dynamic network. \emph{(d)} Small worldness calculations using either the RE (purple) or RP (blue) null model.}
  	\label{fig:4}
  	
  \end{figure}

The degree-preserving configuration model is popular in studies of static graphs because it retains an important aspect of the graph's topology: its degree sequence. However, in a dynamic graph the problem becomes a bit more difficult: we have both edge connectivity and the time dimension which could be randomized. To construct a null model that is most similar to the configuration model for static graphs, we will perform a random rewiring of edges occurring at the same timepoint. More explicitly, for each timepoint $t$ we imagine the static graph $G_t$. We visit each edge of $G_t$ and randomly reassign one end node of this edge to another node within $G_t$, as seen in Fig.~\ref{fig:4}a. We call this the \emph{randomized edges} (RE) model following \cite{holme2012temporal}. Importantly, this null model preserves the contact time component. An alternative is to instead randomize the time at which each contact occurs, giving us the so-called \emph{randomly permuted times} (RP) model (Fig.~\ref{fig:4}b). This model destroys the true temporal contact patterns while preserving overall event rates \cite{holme2012temporal}.

To further illustrate how the RE and RP models alter the temporal structure observed in the original dynamic network, we can calculate the \emph{temporal correlation coefficient} $C = \frac{1}{N}\sum_iC_i$ where 
 
 \begin{equation}
 C_i = \frac{1}{T-1}\sum_{t = 1}^{T-1}\frac{\sum_jA_{i,j}(t)A_{i,j}(t+1)}{\sqrt{[\sum_j A_{i,j}(t)][\sum_j A_{i,j}(t+1)]}}\\,
 \end{equation}
 
\noindent for one subject in the example empirical dynamic graph estimated from fMRI data and for the RP and RE models that were generated from this same graph \cite{tang2010small}. Intuitively, we can think of $C_i$ as the average topological overlap of node $i$'s neighbors between two successive timepoints. As expected, we see in Fig~\ref{fig:4}c that both the RP and RE models have lower values of $C$ than the original dynamic network, indicating that the dynamic graph of the true data is smoothly reconfiguring while the dynamic graphs of the null models are not.

 \subsection*{Temporal small-worldness}

One context in which null models become particularly important is in testing and quantifying the small-worldness of dynamic graphs. Over the last decade, evidence has continued to mount suggesting that neural systems across different species and spatial scales display small-world properties in both structure and function \cite{bassett2006small,bassett2016small}. Yet, little is known about whether or not these systems have \emph{temporal} small-worldness. 

We can recall that the common manner in which one calculates small-worldness for static graphs depends upon estimating the clustering coefficient and the characteristic path length for the original network and appropriate null models \cite{humphries2006brainstem,telesford2011ubiquity,muldoon2016small}. Naturally, if we could generalize each of these to include the time dimension, then we could straightforwardly calculate small-worldness for dynamic graphs as well. First, following \cite{tang2010small} we use the temporal correlation coefficient in place of the clustering coefficient. If a brain region has a high temporal correlation coefficient, then its neighbors persist throughout time in a predictable manner, thereby indicating robust local connections. Next we extend the average shortest path length to temporal networks, giving us the \emph{characteristic temporal path length}, or 
 
 \begin{equation}
 L = \frac{1}{N-1} \sum_{i,j}d(i,j)
 \end{equation}

\noindent where recall $d(i,j)$ refers to the temporal distance (or latency) between two nodes in the network \cite{tang2010small}. 

Now that we have a measure of temporal clustering and of temporal path length, we next turn to the question of whether those values are different than that expected in a random network null model. Specifically, we recall that networks are said to show the small-world property if $\frac{c/c_{rand}}{l/l_{rand}} >1$ where $c_{rand}$ is the static clustering coefficient expected in a random network null model and $l_{rand}$ is the static characteristic path length expected in a random network null model. Extending this notion to dynamic graphs, we can use either the RE or the RP model as the dynamic network null model, and then compute the \emph{temporal small worldness} as $\frac{C/C_{RE}}{L/L_{RE}}$ or $\frac{C/C_{RP}}{L/L_{RP}}$ where $C$ is the temporal correlation coefficient and $L$ is the characteristic temporal path length. In Figure \ref{fig:4}c we apply these notions to the example empirical dynamic graph estimated from fMRI data, and observe that the temporal small-worldness decreases with increasing number of trials practiced.

\subsection*{Temporal Community Structure}

The measures we have discussed thus far have been either focused on individual nodes in the graph or on global, summary statistics of the graph as a whole. Yet an important feature in many networks, particularly in networks representing neurophysiological systems, is mesoscale architecture \cite{newman2006modularity}. Perhaps the most commonly studied type of mesoscale architecture in such networks is community structure \cite{fortunato2010community,fortunato2016community,porter2009communities}: where nodes can be sorted into groups displaying dense intra-group connectivity and sparse inter-group connectivity. Multiple methods for extending community detection to dynamic networks exist \cite{mucha2010community,gauvin2014detecting,ponce2015resting,robinson2015dynamic,betzel2016multi}, and we refer the reader to these resources for more detailed discussions of these methods. Here, we assume that one has applied a dynamic extension of community detection techniques to one's data and has an estimate of each node's affiliation to communities as a function of time. Under these assumptions, we will focus on three metrics that can be used to characterize the fine scale \emph{changes} of communities across time.

\begin{figure*}
 	\centering 
 	
 	\includegraphics[width = .9\linewidth]{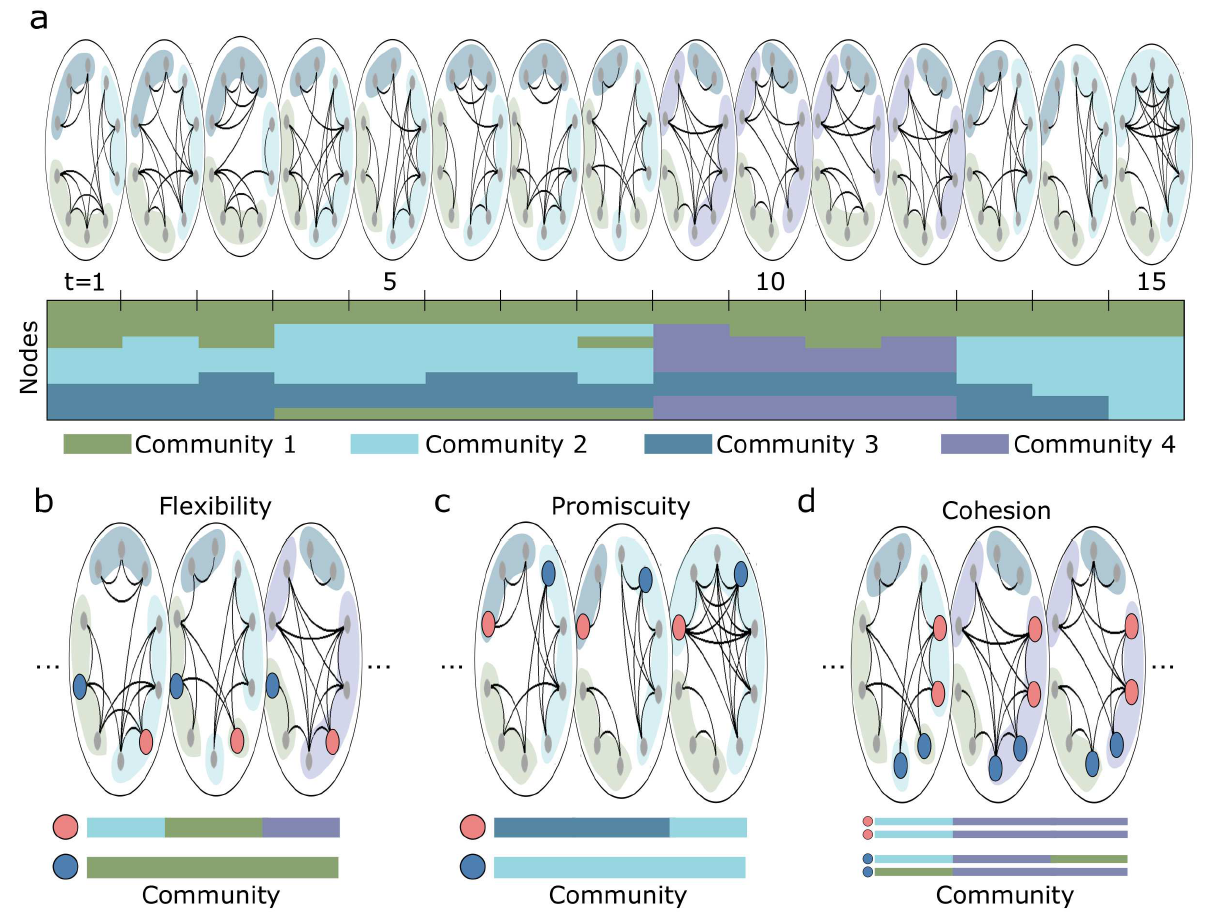}
 	\caption{Metrics associated with dynamic community structure. \emph{(a)} Example dynamic network with a community partition: an assignment of nodes to communities (densely intraconnected groups of nodes) as a function of time. Node community assignments are shown both within a sequence of graphs (top), and as a heatmap (bottom). Examples of nodes with high (orange) and low (blue) values for associated metrics: \emph{(b)} flexibility, \emph{(c)} promiscuity, and \emph{(d)} cohesion.}
 	\label{fig:5}
 	
 \end{figure*}

First, given a community assignment as in Fig.~\ref{fig:5}a, we expect some nodes to likely remain within a single community for all timepoints, while others may change communities often. Within the brain, a node that changes communities multiple times may be modulating multiple processes \cite{fedorenko2014reworking} and may consequentially be essential for dynamic and adaptive processes \cite{bassett2011dynamic}. For example, in the toy dynamic graph displayed in Fig.~\ref{fig:5}b, we see the orange node changes communities three times within the time window, while the blue node remains within the same community. We can quantify this property with the notion of \emph{node flexibility}, defined as the number of times that a node \emph{actually changed} communities, normalized by the number of times the node \emph{could have} changed communities. That is, if node $i$ changed communities $m$ times, the flexibility of node $i$ is 
\begin{equation}
f_i = \frac{m}{T-1}
\end{equation}
\noindent where recall $T$ is the number of timesteps. Then the flexibility $F$ of the dynamic network is the average of $f_i$ over all nodes \cite{bassett2011dynamic}. According to this definition, we see the orange node in Fig.~\ref{fig:5}b has high flexibility, while the blue node has low flexibility. 
 
Yet, simply counting the number of community affiliation swaps for a given node may mask important information. If, for instance, a node of interest swaps back and forth between only two communities, it will have high flexibility but if many other communities exist we cannot infer that it participates in many processes. We see that the node marked in blue in Fig.~\ref{fig:5}c switches between communities 2 and 3 six times throughout the course of the network (Fig.~\ref{fig:5}a, bottom) while the orange node of Fig.~\ref{fig:5}c switches only four times, yet it is at least once a member of all four communities. To better describe this difference we can define \emph{node promiscuity} as 
 
 \begin{equation}
 \psi_i = \frac{k}{K}
 \end{equation}

\noindent for node $i$ which participates in $k$ of $K$ total communities \cite{papadopoulos2016evolution}. Then the promiscuity $\Psi$ of the dynamic network is the average of all $\psi_i$. Intuitively, while flexibility may give one a basic intuition regarding how changeable the community structure is, promiscuity gives one an understanding of how distributed a node's allegiances are to all communities over time.

Since we can measure how nodes change communities across time, we now might ask how groups of nodes change (or do not change) communities. We can assume brain regions that most often change communities in a coordinated fashion are more likely to be involved in the same processes. We define \emph{node cohesion} as the number of times a node changes communities mutually with another node \cite{telesford2017node}. We illustrate this notion pictorially in Fig.~\ref{fig:5}d, where the two orange nodes change communities together, while the blue nodes switch communities independently of each other. In this case, we say the orange nodes are cohesive and the blue nodes are disjoint (or have a low cohesion strength). Using these measures we can probe community dynamics at a finer scale than is possible using community assignments alone.

\section*{Contexts for the Application of Dynamic Graph Metrics}
\label{sec4}

Now that we have described dynamic graph metrics from a mathematical point of view and have illustrated their application to both toy networks and empirical dynamic graphs estimated from fMRI data, we now turn to outlining and discussing a few natural scenarios in which dynamic graphs could be constructed to address hypotheses regarding brain structure and function, as well as the neurophysiological mechanisms of behavior and disease. These scenarios are not meant to be comprehensive, but are simply meant to provide the reader with some intuitions about potential application areas.

\textbf{Cross-scale, Cross-species.}
While we have illustrated these techniques and tools in the context of an fMRI data set, it is important to note that the field of network neuroscience -- which could benefit greatly from dynamic graph tools -- extends far beyond human imaging \cite{bassett2017network}. Arguably even more fundamental are the connectivity patterns characteristic of neuronal circuits, which are measureable, manipulable, and dissectable in non-human animals. This small-scale circuitry displays rich network architectures that can vary over time, development, and species \cite{van2016comparative} and can be explained to some extent by gene co-expression \cite{conaco2012functionalization}. Indeed, prior evidence demonstrates that local cortical circuits display highly nonrandom features of synaptic connectivity \cite{song2005highly,kaiser2006nonoptimal}, characterized by motifs \cite{sporns2004motifs}, distant-dependent architecture \cite{ersey2013predictive}, redundancy \cite{gururangan2014analysis}, and modularity \cite{sadovsky2013scaling}. A particularly interesting set of questions lies in whether and how dynamic graph architectures are conserved across species and to what extent they vary. One might hypothesize that temporal small-worldness -- like static small-worldness -- may be a common design principle across mammalian brains  \cite{bassett2006small,bassett2016small}, arbitrating a dynamic tradeoff between temporal cost and temporal efficiency \cite{bullmore2012}.

\textbf{Cognitive Processes.}
Many cognitive processes are explicitly thought of as dynamic processes, requiring time-dependent changes in information acquisition or retrieval, followed by processing or encoding, to enable responses or decisions. Recent work has demonstrated that functional network architecture in the human brain changes appreciably during such tasks, particularly in those that require higher-order cognitive processing like memory \cite{braun2015dynamic,braun2016dynamic}, attention \cite{shine2016temporal,kucyi2016dynamic}, learning \cite{fatima2016dynamic,heitger2012motor,mantzaris2013dynamic,bassett2014crosslink}, cognitive flexibility \cite{braun2015dynamic}, and task-switching \cite{zalesky2014time}. These types of processes are therefore naturally encoded in dynamic graphs in which the layers of the graph represent time windows, and the edges in the graph represent functional (or effective) connections between neural signals measured from fMRI, EEG, MEG, ECoG, or fNIRS in humans, or calcium transients, local field potentials, etc. in non-human animals. A particularly interesting open question is whether and how these processes are modulated by mood \cite{betzel2017positive} and/or levels of arousal \cite{nassar2012rational}. One might hypothesize that mood instability could manifest as decreases in the temporal correlation coefficient and increases in the temporal path-length, leading to a more random temporal graph. This hypothesis could be tested in future work. 

\textbf{Development and Aging.}
While cognitive processes are accompanied by changes in functional network architecture over relatively short time scales (seconds, minutes, hours), other natural processes evolve over relatively long time scales (months, years, decades). Normal human development and aging are examples of such long-term processes, and recent evidence has begun to map out changes in both structural and functional brain network architecture that track with age \cite{dimartino2014unraveling}. Whether the time frame is fetal development \cite{van2016functional,keunen2017emergence}, child and adolescent development \cite{fair2009functional,gu2015emergence}, or the full lifespan \cite{betzel2014changes,davison2016individual}, patterns of connectivity reconfigure in a manner that at least partially explains changes in cognitive abilities. A particularly striking example is the emergence of cognitive control over development, which has inspired a range of network-based theories pointing towards a critical role for variations in network structure \cite{gu2015controllability,tang2016structural}, network function \cite{marek2015contribution,luna2015integrative}, and network dynamics \cite{hutchison2016matter}. The dynamic graph metrics discussed here offer an interesting and novel framework in which to better probe the relationship between network change and the emergence of cognitive control in fronto-parietal circuitry. In particular, one might hypothesize that the receive centrality of the fronto-parietal network decreases over development, while the broadcast centrality (possibly marking the potential for top-down control) of this same network increases over development. Future work could test this hypothesis explicitly and also test whether the temporal trends in broadcast and receive centrality differ in children with psychosis \cite{satterthwaite2016structural,satterthwaite2015connectome} and those with executive function deficits \cite{shanmugan2016common}.

\textbf{Disease Processes, Disease Progression, Response to Therapy.}
While child-onset psychosis is one condition that may be characterized by altered network dynamics, other neurological disorders and psychiatric disease may also display similar or inherently different sorts of changes \cite{fornito2015connectomics,sharma2017common}. Indeed, recent evidence has demonstrated alterations in the functional network architecture most characteristic of individuals with Alzheimer's disease, Parkinson's disease, and epilepsy to name a few \cite{stam2014modern}.  Interestingly, network architecture can be used to track seizure dynamics \cite{khambhati2016virtual,kramer2011emergence,burns2014network} or the progress of atrophy and dementia \cite{raj2015network}. Less is known about whether and how network architecture or dynamics could be used to track rehabilitation after stroke \cite{ward2004functional} or response to therapeutic interventions including physical therapy \cite{deconinck2015reflections}, brain stimulation \cite{grefkes2012disruption,medaglia2017brain}, and neurofeedback \cite{linden2016realtime,bassett2017neurofeedback,murphy2017network}. Some work suggests that changes in motor behavior are characterized by reconfiguration of functional network modules \cite{bassett2011dynamic,bassett2015learning} and that modularity predicts a person's response to cognitive training after brain injury \cite{arnemann2015functional}. It would be interesting to explicitly test whether the reconfigurations that are most benefitial to stroke rehabilitation are characterized by high flexibility, promiscuity, or cohesion, and whether the relationship between rehabilitation and network reconfiguration is always linear or is better characterized as an inverted U-shaped curve.

\textbf{Extensions to Other Sorts of Graphs.}
While we have focused our exposition on functional dynamic graphs, it is important to note that dynamic graphs can be constructed from many other sorts of data as well. Perhaps the simplest example is a dynamic graph constructed from structural (diffusion imaging tractography) data acquired either over age \cite{betzel2014changes} or training \cite{kahn2016structural}. But one could also consider setting aside the brain entirely and studying network patterns in symptomatology, covariance in markers of mood, or patterns of behavior \cite{wymbs2012differential,acuna2014multifaceted}, where dynamic graphs could provide insight into skill acquisition or adaptive decision-making.

\section*{Conclusion}
\label{sec5}

In summary, we have provided a tutorial on what a dynamic graph actually is, how to visualize it, and how to characterize it. In particular, we describe several basic dynamic graph notions and measures including time-respecting paths, latency, centrality, clustering, characteristic temporal path length, and dynamic modular structure, and we also discuss null models and measures that depend on them, such as temporal small-worldness.  We outline a few natural scenarios in which dynamic graphs could be usefully constructed and studied, and we provide a publically-available MATLAB toolbox to enable the reader to immediately apply these tools to their data.  Our aim is to provide the neuroimaging community with both tools and intuition and to support the growing interest in addressing neuroscientific questions that hinge on detailed analyses of dynamic graphs.

\section*{Acknowledgments}
\label{sec6}

A.E.S. and D.S.B. would like to acknowledge support from the John D. and Catherine T. MacArthur Foundation, the Alfred P. Sloan Foundation, the National Institute of Health (1R01HD086888-01), and the National Science Foundation (BCS-1441502, CAREER PHY-1554488, BCS-1631550). The experiments performed to generate these data were supported by PHS grant NS33494 to Scott T. Grafton; experiments were performed by Nicholas F. Wymbs. The content is solely the responsibility of the authors and does not necessarily represent the official views of any of the funding agencies.

\clearpage
\newpage
\bibliography{bibfile}
\bibliographystyle{plain}

\onecolumn

\section*{Appendix}
\label{appendix}

\begin{tabular}{p{6cm} p{9cm}}
	
	\textbf{Metrics and Definitions}  & \\ 
	\hline \hline &  \\[.5mm]
	
	\textit{Initial definitions} &  Given a dynamic network, we call the vertex set $V$, with $|V| = N$. Edges exist between vertices at any of timepoints $1,2 \dots , T$. May be represented as a sequence of $N \times N$ adjacency matrices $A(1), A(2), \dots, A(T)$. \\[1.5mm]
	
	\textit{Contact} & An edge between two vertices at a specified time.  \\[1.5mm]
	
	\textit{Contact sequence}  & A list of contacts within the dynamic network specified as tuples $(i,j,t)$ for contacts between nodes $i,j$ at time $t$.\\[1.5mm]

	\textit{Time-aggregated graph} & Summary static graph of dynamic network with edges existing between nodes $i,j$ if $i$ and $j$ connect at any timepoint within the dynamic network. \\[1.5mm]
	
	\textit{Time-respecting path} & A sequence of contacts $(n_0, n_1, t_0), (n_1, n_2, t_1) \dots (n_{k-1},n_k, t_{k-1})$ with $t_i<t_{i+1}$ for $i = 0, \dots, k-2$. \\[1.5mm]
	
	\textit{Source set} &  The set of vertices that can reach a given node via time-respecting paths terminating no later than some time $t$. \\[1.5mm]
	
	\textit{Set of influence} & The set of vertices which can be reached from a given node through time-respecting paths starting no earlier than some time $t$. \\[1.5mm]
	
	\textit{Temporal path length} &  The difference in time between the last and first contact of a time-respecting path \cite{pan2011path}.\\[1.5mm]
	
	\textit{Latency} & The temporal path length of the fastest path between two nodes. Also known as temporal distance \cite{pan2011path}.\\[1.5mm]
	
	\textit{Forward Latency} & Denoted $\tau(i,j,t)$, the time needed to reach node $j$ from $i$ along time-respecting paths beginning no earlier than $t$ \cite{pan2011path}.\\[1.5mm]
	
	\textit{Betweenness centrality} & For node $i$ and timepoint $t$, $$C_B(i,t) = \sum_{i\neq j \neq k}\frac{\sigma_{j,k}(i,t)}{\sigma_{j,k}(t)}$$ with $\sigma_{j,k}$ the number of shortest paths between nodes $j$, $k$ beginning no earlier than $t$, and $\sigma{j,k}{t,i}$ the number of such paths that pass through node $i$ \cite{tang2010analysing}. \\[1.5mm]
	
	\textit{Closeness centrality} & For node $i$ and time $t$, $$C_C(i,t) = \frac{1}{N-1}\sum_{j \neq i} \frac{1}{\tau(i,j,t)}$$ \cite{pan2011path}. \\[1.5mm]
	
	\textit{Broadcast centrality} &  Given node $i$, the broadcast centrality is $$b(i) := \sum_{j=1}^N Q_{i,j}$$ where $Q_{i,j}$ is the normalized ability of node $i$ to communicate with node $j$ (See Eq. \ref{eq:broadcast2}) \cite{mantzaris2013dynamic}.\\[1.5mm]

\end{tabular}

\newpage
\begin{tabular}{p{5.5cm} p{10cm}}
		\textit{Receive centrality} & Given node $j$, receive centrality is defined $$r(j) := \sum_{i=1}^N Q_{i,j} \hspace{2pc}$$ \cite{mantzaris2013dynamic}. \\[1.5mm]
		
		\textit{Temporal correlation coefficient} & Let $A_{i,j}(t)$ be the connectivity of nodes $i$, $j$ at time $T$. Then for node $i$, $$ C_i = \frac{1}{T-1}\sum_{t = 1}^{T-1}\frac{\sum_jA_{i,j}(t)A_{i,j}(t+1)}{\sqrt{[\sum_j A_{i,j}(t)][\sum_j A_{i,j}(t+1)]}}$$ \cite{tang2010small}. \\[1.5mm]
		
	\textit{Characteristic temporal path length} &  For a dynamic network, $$L = \frac{1}{N-1} \sum_{i,j}d(i,j)$$ letting $d(i,j)$ be the temporal distance between nodes $i$, $j$.\\[1.5mm]
	
	\textit{Temporal small worldness} & Let $C$, $C_{rand}$ be the average temporal correlation coefficient and $L$, $L_{rand}$ the temporal characteristic path length for the dynamic network and randomized model, respectively. Then the temporal small worldness is $$\frac{C/C_{rand}}{L/L_{rand}}$$ \cite{tang2010small}. \\[1.5mm]
	
	\textit{Flexibility} & For node $i$, the flexibility is $$f_i = \frac{m}{T-1}$$ where $m$ is the number of times node $i$ change communities \cite{bassett2011dynamic}.\\[1.5mm]
	
	\textit{Promiscuity} & The promiscuity of node $i$ is $$\psi_i = \frac{k}{K}$$ with $k$ the number of communities of which node $i$ is a member and $K$ the total number of communities in the dynamic network \cite{papadopoulos2016evolution}.\\[1.5mm]
	
	\textit{Cohesiveness} & The number of times a node changes communities mutually with another node \cite{telesford2017node}.

\end{tabular}

\end{document}